# Solar-system tests of the relativistic gravity*§

Wei-Tou Ni

*School of Optical-Electrical and Computer Engineering,
University of Shanghai for Science and Technology,
516, Jun Gong Rd., Shanghai 200093, China* weitou@gmail.com

In 1859, Le Verrier discovered the Mercury perihelion advance anomaly. This anomaly turned out to be the first relativistic-gravity effect observed. During the 157 years to 2016, the precisions and accuracies of laboratory and space experiments, and of astrophysical and cosmological observations on relativistic gravity have been improved by 3-4 orders of magnitude. The improvements have been mainly from optical observations at first followed by radio observations. The achievements for the past 50 years are from radio Doppler tracking and radio ranging together with lunar laser ranging. At the present, the radio observations and lunar laser ranging experiments are similar in the accuracy of testing relativistic gravity. We review and summarize the present status of solar-system tests of relativistic gravity. With planetary laser ranging, spacecraft laser ranging and interferometric laser ranging (laser Doppler ranging) together with the development of drag-free technology, the optical observations will improve the accuracies by another 3-4 orders of magnitude in both the equivalence principle tests and solar-system dynamics tests of relativistic gravity. Clock tests and atomic interferometry tests of relativistic gravity will reach an ever-increasing precision. These will give crucial clues in both experimental and theoretical aspects of gravity, and may lead to answers to some profound issues in gravity and cosmology.

*Keywords:* General relativity, Experimental tests of relativistic gravity, Solar-system dynamics, Ephemerides

PACS Numbers: 04.80.Cc, 04.80.-y, 95.10.-a

## 1. Introduction and Summary

The development of gravity theory stems from experiments. Newton's theory of gravity [1] is empirically based on Kepler's laws [2] (which are based on Brahe's observations) and Galileo's law of free-falls [3] (which is based on Galileo's experiment of motions on inclined planes). Toward the middle of nineteenth century, astronomical observations accumulated a precision which enabled Le Verrier [4] to discover the Mercury perihelion advance anomaly in 1859. This anomaly is the first relativistic-gravity effect observed. Michelson-Morley experiment [5], via various developments [6], prompted the final establishment of the special relativity theory in 1905 [7, 8]. Motivation for putting electromagnetism and gravity into the same theoretical framework [7], the precision of Eötvös experiment [9] on the equivalence, the formulation of Einstein equivalence principle [10] together with the perihelion advance anomaly led to the road for the final genesis of general relativity (GR) theory [11-15] in 1915.

---

*This review is dedicated to my mother Suh-Ling Huang (1923.12.13 – 2016.01.24) on her obituary 2016.01.29. Her amity, competence and diligence as mother fostered my confidence and perseverance to continue.

§This paper is to be published also in the book "*One Hundred Years of General Relativity*: *From Genesis and Empirical Foundations to Gravitational Waves, Cosmology and Quantum Gravity*", edited by Wei-Tou Ni (World Scientific, Singapore, 2016).

As we discussed in [16, 17], Einstein proposed the mass-energy equivalence using the formula $E = mc^2$ in 1905 [18]; Planck reasoned that all energy must gravitate in 1907 [19]. To characterizing the strength of a gravitational source, it would then be natural to compare magnitude of the gravitational energy $mU(x, t)$ of a test particle in the gravitational potential $U(x, t)$ of a gravitating source to the total mass-energy $mc^2$ of the test particle and define this ratio $\xi(x, t)$ as the dimensionless gravitational strength of the source at a spacetime point $\wp$ [with coordinates $(x, t)$]:

$$\xi(x, t) = U(x, t) / c^2. \tag{1}$$

General Relativity (GR) gives strong-field corrections to the Newtonian gravity. The first order correction is proportional to this strength $\xi(x, t)$.

For a point source with mass $M$ in Newtonian gravity,

$$\xi(x, t) = GM / Rc^2, \tag{2}$$

where $R$ is the distance to the source. For a nearly Newtonian system, we can use Newtonian potential for $U$. The strength of gravity for various configurations is tabulated in Table 1.

Table 1. The strength of gravity for various configurations.

| Source | Field Position | Strength of Gravity $\xi$ |
|---|---|---|
| Sun | Solar Surface | $2.1 \times 10^{-6}$ |
| Sun | Mercury Orbit | $2.5 \times 10^{-8}$ |
| Sun | Earth Orbit | $1.0 \times 10^{-8}$ |
| Sun | Jupiter Orbit | $1.9 \times 10^{-9}$ |
| Earth | Earth Surface | $0.7 \times 10^{-9}$ |
| Earth | Moon's Orbit | $1.2 \times 10^{-11}$ |
| Galaxy | Solar System | $10^{-5} - 10^{-6}$ |
| Significant Part of Observed Universe | Our Galaxy | $1 - 10^{-2}$ |

From Table 1, it is clear that in the solar system Mercury has the largest solar-system gravitational potential among all planets and satellites, and hence the largest general-relativistic solar-system gravitational correction. This is why the general-relativistic deviation of the Mercury orbit from Newtonian theory – the Mercury perihelion advance anomaly of about 40" per century was first observed. When the observations reached an accuracy of the order of 1" per century (transit observations) in the 19th century, a discrepancy from Newtonian gravity would be seen. In a century, Mercury orbits around the Sun 400 times, amounting to a total angle of $5 \times 10^8$ arc sec. The fractional relativistic correction (perihelion advance anomaly) of Mercury's orbit is of order $\lambda GM_{Sun}/dc^2$ with $\lambda = 3$ for GR (i.e. about $8 \times 10^{-8}$) and $d$ the distance of Mercury to the Sun. Therefore, the relativistic correction for perihelion advance is about 40 arc sec per century. As the orbit determination of Mercury reached an accuracy better than $10^{-8}$ (about 1 arc sec for solar transit observations in 100 years), the relativistic corrections to Newtonian gravity became manifest. Le Verrier discovered this perihelion advance anomaly (anomaly to Newton's theory) and measured it to be 38 arc sec per century [4]. In 1881 Newcomb obtained a more precise value (43 arcsecond per century) of Mercury perihelion advance anomaly [20].

In 1907 Einstein proposed his equivalence principle and derived the gravitational redshift [10]; in 1911 Einstein derived the light deflection in the solar gravitational field [21].[a] In 1913 Besso and Einstein [27] worked out a Mercury perihelion advance formula in the "Einstein-Grossmann Entwert" theory [28], but the calculation contained an error and did not agree with the experimental value. During the final genesis of general relativity [11-15], Einstein [13] corrected their 1913 error and obtained a Mercury perihelion advance value in agreement with the observation [20]. Apparently, this correct calculation played a significant role in the final genesis of general relativity.

Gravitational redshift, gravitational light deflection and relativistic perihelion advance are called 3 classical tests of general relativity in Einstein's "The Foundation of the General Theory of Relativity." [29]

Toward the end of the nineteenth century, there were studies whether the solar spectra were displaced from the Doppler spectra [30, 31]. Various causes (such as pressure effect, pole effect, asymmetrical broadening) were found and investigated before 1910 [32-35]. In 1911, Einstein [21] noticed the work of Buisson and Fabry [34], and explicitly proposed that gravitational redshift might be tested by the examination of the solar spectra. From 1914 to 1919, re-analysis of previous solar spectra together with a number of new measurements were made. However, the outcome is controversial and inconclusive. Earman and Glymour [36] gives a detailed account of this history.

Before Einstein's proposal of relativistic solar deflection of light in 1911, there were photographs taken for studying the solar corona and to find a sub-Mercurial planet of solar neighborhood during total eclipses. These photographs were considered unsatisfactory to study the deflection of light by Perrine upon a question from Freundlich late 1911 either because of small field and brief exposure time, or because of eccentric position of the Sun on the plates (See, e.g. p. 61 of [37]). Before 1919, there were 4 expeditions intent to measure the gravitational deflection of starlight (in 1912, 1914, 1916 and 1918); because of bad weather or war, the first 3 expeditions failed to obtain any results, the results of 1918 expedition was never published [37]. In 1919, the observation of gravitational deflection of light passing near the Sun during a solar eclipse [38] confirmed the relativistic deflection of light and made general relativity famous and popular.

___________

[a]Newton in his Opticks [22] of 1704 proposed the following query for further research: "Do not Bodies act upon Light at a distance, and by their action bend its Rays, and is not this action strongest at the least distance?" In 1801, Soldner [23] derived the gravitational bending of light from corpuscular nature of light and Newton's universal gravitation. Soldner [23, 24] calculated the deflection angle for light grazing the Sun to be 0.84 arcsec remarkably close to Einstein's 1911 value of 0.83 arcsec [21]. Cavendish's work on the gravitational bending of light (probably around 1784 [25]) was published posthumously in 1921 [26].

The success of Pound and Rebka [39] in using Mössbauer effect to verify the gravitational redshift in earth-bound laboratory in 1960 marked the beginning of a new era for testing relativistic gravity. At the same time, a careful and more precise test of the equivalence principle was performed in Princeton [40]. With the development of technology and advent of space era, Shapiro [41] proposed a fourth test --- the time delay of radar echoes in gravitational field. Since the beginning of this era, we have seen 3-4 orders of improvements for the three classical tests together with many new tests. The current technological development is ripe that we are now in a position to discern another 3-4 orders of improvements further in testing relativistic gravity in the coming 25 years (2016-2040). This will enable us to test the second order relativistic-gravity effects. A road map of experimental progress in gravity together with its theoretical implication has been shown in Table 2 of Ref. [42].

The present review updates the solar-system test part of a previous review on "Empirical Foundations of the Relativistic Gravity" [42] (which is a five-year update of the 1999-2000 review [43]). A companion review on equivalence principles and the foundation of metric theories of gravity has been already given in [17]. Recently Manchester [44] has reviewed the pulsar tests of relativistic gravity. A previous review on the solar-system tests of relativistic gravity is Reynaud and Jaekel [45]. A good general review on experimental tests of general relativity is Will [46].

In section 2, we review the post-Newtonian approximation of General Relativity, the PPN (Parametrized Post-Newtonian) framework, and derive the Shapiro time delay and the first order relativistic light deflection as examples. In section 3, we review and discuss the solar-system ephemerides. In section 4, we update the solar system tests since our last review in 2005. In section 5, we discuss ongoing and next generation solar-system experiments related to testing relativistic gravity with an outlook.

## 2. Post-Newtonian Approximation, PPN Framework, Shapiro Time Delay and Light deflection

The equations of motion of General Relativity, i.e. the Einstein equation is

$$G_{\mu\nu} = \kappa\, T_{\mu\nu}, \tag{3}$$

where $T_{\mu\nu}$ is the stress-energy tensor and $\kappa = 8\pi G/c^4$ (See, e.g. [47]). We use the MTW[47] conventions with signature $-2$; This is also the conventions used in [16, 17]; Greek indices run from 0 to 3; Latin indices run from 1 to 3; the cosmological constant is negligible for solar-system dynamics and solar-system ephemeris, and is neglected in this review. Contracting the equations of motion (3), we have

$$R = -(8\pi/c^4)\, G\, T, \tag{4}$$

where $T \equiv T_\mu^{\ \mu}$. Substituting (4) into (3), we obtain the following equivalent equations of motion as originally proposed by Einstein [15]

$$R_{\mu\nu} = (8\pi/c^4)\, [T_{\mu\nu} - (1/2)(g_{\mu\nu}T)]. \tag{5}$$

For weak field in the quasi-Minkowskian coordinates, we express the metric $g_{\alpha\beta}$ as

$$g_{\alpha\beta} = \eta_{\alpha\beta} + h_{\alpha\beta}, \quad h_{\alpha\beta} \ll 1. \tag{6}$$

Since $h_{\alpha\beta}$ is a small quantity ($< 4 \times 10^{-6}$) in the solar-system gravitational field, we expand everything in $h_{\alpha\beta}$ and linearize the results to obtain the linear (weak-field) approximation. With the harmonic gauge (coordinate) condition for $h_{\alpha\beta}$,

$$[h_{\alpha\beta} - (1/2)\eta_{\alpha\beta}(\text{Tr } h)]^{,\beta} = 0 + O(h^2), \text{ i.e., } h_{\alpha\beta}{}^{,\beta} = (1/2)(\text{Tr } h)_{,\alpha} + O(h^2), \tag{7}$$

the linearized Einstein equation is:

$$h_{\mu\nu,\beta}{}^{\beta} = -(16\pi G/c^4)[T_{\mu\nu} - (1/2)(\eta_{\mu\nu}T)] + O(h^2), \tag{8}$$

where Tr($h$) is defined as the trace of $h_\alpha{}^\beta$, i.e. Tr($h$) $\equiv h_\alpha{}^\alpha$, and $O(h^2)$ denotes terms of order of $h_{\alpha\beta}h_{\mu\nu}$ or smaller (See, e.g. [47,48]). Analogous to classical electrodynamics, the solution of this equation for GR is

$$h_{\mu\nu} = -[(4G)/(c^4)] \int \{[T_{\mu\nu} - (1/2)g_{\mu\nu}T]/r\}_{\text{retarded}} (d^3x') + O(h^2). \tag{9}$$

### 2.1. Post-Newtonian approximation

For solar dynamics and solar-system ephemerides, we can impose slow motion condition, in addition to weak field condition, i.e.

$$U/c^2 = O(v^2/c^2); \ U_{,ij}/c^2 = (1/L^2) O(v^2/c^2); \ U_{,0i}/c^2 = (1/L^2) O(v^3/c^3); \ U_{,00}/c^2 = (1/L^2) O(v^4/c^4), \tag{10}$$

where $L$ is a typical length scale and $v$ is a typical velocity of the system (See, e.g., Ref. [47]). The solution $h_{\mu\nu}$ of (9) and $h_\alpha{}^\alpha$ in this approximation then becomes

$$h_{\mu\nu} = -2(U/c^2)\delta_{\mu\nu} + O(v^3/c^3); \ h \equiv h_\mu{}^\mu = 4(U/c^2) + O(v^3/c^3), \tag{11}$$

where $U$ is the Newtonian potential which normally contains multipole terms outside a gravitating body. For point mass or outside spherical Sun,

$$U = (GM/c^2)(1/r); \ r = (x^2 + y^2 + z^2)^{1/2}. \tag{12}$$

With the metric (11), one can already derive the solar deflection of light and the Shapiro time delay. For a derivation of relativistic precession of Mercury's orbit, one needs a full post-Newtonian approximation of GR and needs to calculate $h_{00}$ to $O(v^4/c^4)$ order and $h_{0i}$ to $O(v^3/c^3)$ order. The post-Newtonian approximation for perfect fluid in general relativity is obtained by Chandrasekhar [49]. The metric $g_{\alpha\beta}$ ($= \eta_{\alpha\beta} + h_{\alpha\beta}$) is given by

$$g_{00} = 1 - 2U/c^2 + 2U^2/c^4 + 4\Psi + O(v^5/c^5),$$
$$g_{0i} = (7/2)V_i + (1/2)W_i + O(v^5/c^5),$$
$$g_{ij} = -(1 + 2U/c^2)\delta_{ij} + O(v^4/c^4), \tag{13}$$

where

$$U(\mathbf{x}, t) = \int [\rho_0(\mathbf{x}', t) / |\mathbf{x} - \mathbf{x}'|] \, dx', \tag{14}$$

$$\Psi(\mathbf{x}, t) = \int [\rho_0(\mathbf{x}', t) \, \psi(\mathbf{x}', t) / |\mathbf{x} - \mathbf{x}'|] \, dx', \tag{15}$$

$$\psi = \mathbf{v}^2 + U + (1/2) \, \Pi + (3/2) \, p/\rho_0, \tag{15a}$$

$$V_i(\mathbf{x}, t) = \int [\rho_0(\mathbf{x}', t) \, v_i(\mathbf{x}', t) / |\mathbf{x} - \mathbf{x}'|] \, dx', \tag{16}$$

$$W_i(\mathbf{x}, t) = \int \{\rho_0(\mathbf{x}', t) \, [(\mathbf{x} - \mathbf{x}') \cdot \mathbf{v}(\mathbf{x}', t)] / |\mathbf{x} - \mathbf{x}'|^3\} \, dx', \tag{17}$$

with $\rho_0(\mathbf{x}, t)$ the rest mass density, $\mathbf{v}(\mathbf{x}, t)$ [= $(v_1, v_2, v_3)$] the 3-velocity, $U(\mathbf{x}, t)$ the Newtonian potential, $\Pi(\mathbf{x}, t)$ the internal energy and $p(\mathbf{x}, t)$ the pressure of the fluid.

*2.2. Parametrized Post-Newtonian (PPN) framework*

For different theories of relativistic gravity, the post-Newtonian metrics are different. However, the post-Newtonian metric of many relativistic gravity theories can be encompassed in the PPN (Parametrized Post-Newtonian) framework with nine post-Newtonian parameters $\beta$, $\beta_1$, $\beta_2$, $\beta_3$, $\beta_4$, $\gamma$, $\zeta$, $\Delta_1$ and $\Delta_2$ [50-54]:

$$g_{00} = 1 - 2 \, U/c^2 + 2 \, \beta \, U^2/c^4 - 4 \, \underline{\Psi} + \zeta \, \mathcal{Q} + O(v^5/c^5),$$

$$g_{0i} = (7/2) \, \Delta_1 \, V_i + (1/2) \, \Delta_2 \, W_i + O(v^5/c^5),$$

$$g_{ij} = -(1 + 2 \, \gamma \, U/c^2) \, \delta_{ij} + O(v^4/c^4), \tag{18}$$

where

$$\underline{\Psi}(\mathbf{x}, t) = \int [\rho_0(\mathbf{x}', t) \, \underline{\psi}(\mathbf{x}', t) / |\mathbf{x} - \mathbf{x}'|] \, dx', \tag{19}$$

$$\underline{\psi} = \beta_1 \mathbf{v}^2 + \beta_2 \, U + (1/2) \, \beta_3 \, \Pi + (3/2) \, \beta_4 \, p/\rho_0, \tag{20}$$

$$\mathcal{Q}(\mathbf{x}, t) = \int \{\rho_0(\mathbf{x}', t) \, [(\mathbf{x} - \mathbf{x}') \cdot \mathbf{v}(\mathbf{x}', t)]^2 / |\mathbf{x} - \mathbf{x}'|^3\} \, dx'. \tag{21}$$

Each gravity theory has a specific set of values for these PPN parameters if it can be encompassed in the framework. GR has the PPN parameters $\beta = \gamma = 1$, $\beta_1 = \beta_2 = \beta_3 = \beta_4 = \Delta_1 = \Delta_2 = 1$, and $\zeta = 0$. Brans-Dicke-Jordan theory has the PPN parameters $\beta = 1$, $\gamma = (1 + \omega) / (2 + \omega)$, $\beta_1 = (3 + 2\omega) / (4 + 2\omega)$, $\beta_2 = (1 + 2\omega) / (4 + 2\omega)$, $\beta_3 = 1$, $\beta_4 = (1 + \omega) / (2 + \omega)$, $\zeta = 0$, $\Delta_1 = (10 + 7\omega) / (14 + 7\omega)$ and $\Delta_2 = 1$ with $\omega$ the Brans-Dicke parameter. Brans-Dicke-Jordan theory is a scalar-tensor theory. For general scalar-tensor theories without mass terms, their PPN parameters are $\beta = 1 + \Lambda$, $\gamma = (1 + \omega) / (2 + \omega)$, $\beta_1 = (3 + 2\omega) / (4 + 2\omega)$, $\beta_2 = (1 + 2\omega) / (4 + 2\omega) - \Lambda$, $\beta_3 = 1$, $\beta_4 = (1 + \omega) / (2 + \omega)$, $\zeta = 0$, $\Delta_1 = (10$

+ 7ω) / (14 + 7ω) and $Λ_2 = 1$ with $Λ$ a second parameter in addition to $ω$. Both general relativity and scalar-tensor theories are conservative and non-preferred-frame theories. For them, it would be more convenient to re-define the following linear combinations of parameters as the new PPN parameters [55, 56]:

$$\alpha_1 \equiv 7 Λ_1 + Λ_2 - 4 γ - 4,$$
$$\alpha_2 \equiv Λ_2 + ζ - 1,$$
$$\alpha_3 \equiv 4 β_1 - 2 γ - 2 - ζ,$$
$$ζ_1 \equiv ζ,$$
$$ζ_2 \equiv 2 β + 2 β_2 - 3 γ - 1,$$
$$ζ_3 \equiv β_3 - 1,$$
$$ζ_4 \equiv β_4 - γ. \tag{22}$$

In terms of the 9 parameters $β, γ, α_1, α_2, α_3, ζ_1, ζ_2, ζ_3$, and $ζ_4$, the only parameters which are not vanishing for GR and for scalar-tensor theories are the two parameters $β$ and $γ$. Indeed, Will and Nordtvedt [55, 56] showed that a theory of gravity which can be encompassed in the PPN framework at the post-Newtonian level and which possesses all ten global conservation laws (four for energy-momentum and six for angular momentum) if and only if

$$ζ_1 = ζ_2 = ζ_3 = ζ_4 = α_3 = 0. \tag{23}$$

$α_1, α_2$, and $α_3$ measure the extent and nature of preferred-frame effects [53, 56, 47]; any gravity theory with at least one of $α_i$'s nonzero is called a preferred-frame theory. In the PPN framework (18), conservative non-preferred-frame theories can have only two independent parameters $β$ and $γ$. General scalar-tensor theories without mass term span the whole class of conservative theories fitted in the PPN framework (18).

Empirically, the preferred-frame and non-conservative parameters $α_1, α_2, α_3, ζ_1, ζ_2$, and $ζ_3$ are constrained as follows:

$|α_1| < 3.4 \times 10^{-5}$ (limit [57] from the orbit dynamics of the binary pulsar PSR J1738 + 0333 [58]),

$|α_2| < 1.6 \times 10^{-9}$ (limit from millisecond pulsars PSR B1937 +21 and PSR J1744 − 1134 [59]),

$|α_3| < 4.0 \times 10^{-20}$ (limit from the orbital dynamics of the statistical combination of a set of binary pulsars [60]),

|ζ$_1$| < 1.5 ×10$^{-3}$ (limit calculated from the constraints on the Nordtvedt parameter η [= 4β − γ − 3 − α$_1$ + (2/3) α$_2$ − (2/3) ζ$_1$ − (1/3) ζ$_2$] and other parameters [Table 2]),

|ζ$_2$| < 4 × 10$^{-5}$ (limit from binary pulsar PSR 1913+16 acceleration [61]),

|ζ$_3$| < 1.5 ×10$^{-3}$ (limit from confirmation of Newton's third law by lunar acceleration [62-64]), (24)

As to ζ$_4$, according to Will [46, 65], there is a theoretical relation 6ζ$_4$ = 3α$_3$ + 2ζ$_1$ − 3ζ$_3$ for gravity theories whose perfect-fluid equations are blind to different forms of internal energy and pressure in the fluid so that ζ$_4$ becomes redundant.

Although PPN framework (18) encompasses a large class of gravity theories, there are still many gravity theories outside its scope. One notable example is Whitehead theory as completed by imposing the Einstein Equivalence Principle. Its post-Newtonian metric contains additional terms which has to be parametrized by an additional parameter ξ (or ξ$_W$) called Whitehead parameter. These additional terms with parameter ξ can be included in an extended PPN framework [55, 56]. For Whitehead theory, ξ = 1 (by definition). Solar-system tests and constraints on the Whitehead terms have been studied in [66-69] with |ξ| constrained to order of 10$^{-3}$. The constraint from millisecond pulsars [70] gives |ξ| < 3.9 × 10$^{-9}$. Also, the PPN framework (18) does not contain the intermediate-range gravity terms (Yukawa terms). These terms can be included in a separate treatment. Misner, Thorne and Wheeler [47] have treated the case with the anisotropic stresses. For this case, there is a post-Newtonian term in g$_{00}$ with an extra parameter in addition to parameter β$_4$. However, the anisotropic stresses are much smaller than the isotropic stresses or pressures in the solar system. For solar-system dynamics consideration, they are negligible up to now.

Historically, Eddington [71] first used the parametrization of metric for discussing the classical tests of relativistic gravity based on the isotropic post-Newtonian expansion of the Schwarzschild metric with the following line element:

$$ds^2 = [1−2α(GM/r)+2β(GM/r)^2+…]dt^2 − [1+2γ(GM/r)+…] (dr^2+r^2dθ^2+r^2\sin^2θdφ^2),$$ (25)

where α, β, γ are called the Eddington parameters. For metric theories, Einstein Equivalence Principle (EEP) is assumed already. The Eddington parameters should not depend on the mass-energy contents. To have the correct Newtonian limit, α can be absorbed into a re-definition of Newtonian gravitational constant G (i.e., αG → G). Hence as a parametrized post-Newtonian framework, there are only 2 effective Eddington parameters β and γ. In 1968, Nordtvedt [72] developed the first modern version of PPN framework for a system of two gravitating point masses with later generalization to more particles. It contains seven parameters in addition to α. In 1971, Will [52] extended the PPN framework to perfect fluid with 2 additional parameters β$_3$ and β$_4$ (or ζ$_3$ and ζ$_4$ in the new combination of parameters) for terms on internal energy and pressure. This framework does not contain the parameter $\alpha$. As we noticed in the comment after (24), $\zeta_4$ (or $\beta_4$) would be a redundant parameter if EEP is assumed. So is $\zeta_3$ (or $\beta_3$). They are really parameters for testing EEP. As we discussed at the beginning of this paragraph, $\alpha$ parameter tests EEP too. In a metric framework like PPN framework (18), seven parameters can be explored. This means the 1968-Nordtvedt and 1971-Will framework are effectively equivalent.

Moreover, the preferred-frame parameters and the conservation-law parameters $\alpha_1$, $\alpha_2$, $\zeta_1$, $\zeta_2$, and $\alpha_3$ are essentially test parameters for strong equivalence principles. They are constrained to observe strong equivalence principles quite well [See e.g., (24)]. In the rest of this article we concentrate mainly on the experiments to test the two parameters $\beta$ and $\gamma$. In the next two subsections, we illustrate the PPN effects with two simple calculations of Shapiro time delay and the gravitational deflection of light passing near the Sun.

*2.3. Shapiro time delay*

One can derive the light propagation equation in the weak field limit for the physical metric (6). Let $\underline{r} = \underline{r}(t)$ be the light trajectory where $\underline{r}(t) = (x(t), y(t), z(t))$ is a 3-vector. Light propagation follows null geodesics of the metric $g_{\alpha\beta}$; its trajectory $\underline{r}(t)$ satisfies

$$0 = ds^2 = g_{\alpha\beta}dx^\alpha dx^\beta = (1 + h_{00})c^2 dt^2 + 2h_{0i}cdx^i dt + (\eta_{ij} + h_{ij})dx^i dx^j, \qquad (26)$$

using (6).

In the Minkowski approximation, the light trajectory can be approximated by

$$dx^i/dt = (dx^i/dt)^{(0)i} + O(h) = cn^{(0)i} + O(h), \text{ with } \sum_i (n^{(0)i})^2 = 1, \qquad (27)$$

where $n^{(0)i}$ are constants. In the Post-Minkowski approximation, we express $dx^i/dt$ as

$$dx^i/dt = cn^{(0)i} + cn^{(1)i} + O(h^2), \qquad (28)$$

where $n^{(1)i}$ is a function of trajectory and of the order of O($h$). Substituting (28) into (26) and dividing by $dt^2$, we have

$$0 = (1 + h_{00})c^2 + 2h_{0i}c(dx^i/dt) + |d\underline{r}/dt|^2 - h_{ij}[(dx^i/dt)(dx^j/dt)] \qquad (29a)$$
$$= (1 + h_{00})c^2 + 2h_{0i}c(cn^{(0)i} + cn^{(1)i}) - c^2\sum_{i=1}^3 (n^{(0)i} + n^{(1)i})^2 + h_{ij}n^{(0)i}n^{(0)j}c^2 + O(h^2). \qquad (29b)$$

Simplifying (29b), we have

$$\sum_{i=1}^3 n^{(0)i}n^{(1)i} = (1/2)(h_{00} + 2h_{0i}n^{(0)i} + h_{ij}n^{(0)i}n^{(0)j}) + O(h^2), \qquad (30)$$

and solving for $|d\underline{r}/dt|$ in (29a), we obtain the light propagation equation to O($h$) order:

$$|d\underline{r}/dt| = c[(1 + h_{00} + 2h_{0i}n^{(0)i} + h_{ij}n^{(0)i}n^{(0)j} + O(h^2)]^{1/2}$$
$$= c[1 + (1/2)h_{00} + h_{0i}n^{(0)i} + (1/2)h_{ij}n^{(0)i}n^{(0)j} + O(h^2)]. \qquad (31)$$

From (31), we calculate the light travel time $\Delta t_{TT}$ between two observers (time delay) [42] as

$$\Delta t_{TT} = (1/c)\int |d\,\mathbf{r}\,|\,[1 - (1/2)h_{00} - h_{0i}n^{(0)i} - (1/2)\,h_{ij}n^{(0)i}n^{(0)j} + O(h^2)]. \tag{32}$$

Choosing the z-axis along the initial light propagation direction, i.e., $n^{(0)i} = (0, 0, 1)$, and using (18) or (25) for a slow-motion observer and Sun (or a central mass), we have

$$\begin{aligned}\Delta t_{TT} &= \int dt = (1/c)\int dz[1 + (1 + \gamma)\,U + O(h^2)] = \Delta t^N + [(1 + \gamma)/2]\,\Delta t_S^{GR}\\ &= (1/c)\,(z_2 - z_1) + (1 + \gamma)\,(GM/c^3)\,ln\{[(z_2^2 + b^2)^{1/2} + z_2]/[(z_1^2 + b^2)^{1/2} + z_1]\} + O(h^2),\\ &\quad (z_1 < 0,\ z_2 > 0)\end{aligned} \tag{33}$$

where the first term is the Newtonian travel time $\Delta t^N$ (Römer delay), the second term is the relativistic Shapiro time delay [41] with $\Delta t_S^{GR}$ the general relativistic Shapiro time delay, and $b$ is the impact parameter of light propagation to the Sun.

*2.4. Light deflection*

The geodesic equation for light and for test particle in general relativity and in the metric theories of gravity

$$d^2x^\mu/d\lambda^2 + \Gamma^\mu_{\sigma\rho}(dx^\sigma/d\lambda)(dx^\rho/d\lambda) = 0, \qquad \lambda: \text{affine parameter} \tag{34}$$

can be cast in the form

$$d(g_{\mu\nu}dx^\nu/d\lambda)/d\lambda = (1/2)\,g_{\sigma\rho,\mu}(dx^\sigma/d\lambda)(dx^\rho/d\lambda). \tag{35}$$

Integrating, we obtain

$$(g_{\mu\nu}dx^\nu/d\lambda)|_{x0}^{x1} = (1/2)\,\int_{x0}^{x1}\,[g_{\sigma\rho,\mu}(dx^\sigma/d\lambda)(dx^\rho/d\lambda)]\,d\lambda. \tag{36}$$

To obtain light deflection angle in a weak gravitational field of the Sun or other source, we choose x-axis in the initial light (photon) propagation direction, y-axis in the plane spanned by the Sun or other gravitational source and the light trajectory, and the sense of the x-axis is in the direction of the trajectory. From the $\mu = y$ component of (36), we obtain

$$(g_{0y} + g_{xy} - (1/c)\,dy/dt)|_{x0}^{x1} = (1/2)\,\int_{x0}^{x1}\,(h_{00,y} + h_{xx,y} + 2h_{0x,y})\,cdt + O(h^2). \tag{37}$$

Solving for $dy/dt$ in (37), substituting (18) or (25) in and simplifying, we obtain

$$\begin{aligned}\Delta\varphi_{deflection} &= (1/c)\,(dy/dt)|_{x0}^{x1} = (g_{0y}+g_{xy})|_{x0}^{x1} - (1/2)\int_{x0}^{x1}(h_{00,y}+h_{xx,y}+2h_{0x,y})\,cdt + O(h^2)\\ &= \{-[(1+\gamma)/c^2b]\,GMx/(x^2+b^2)^{1/2}\}|_{x=x1} + \{[(1+\gamma)/c^2b]\,GMx/(x^2+b^2)^{1/2}\}|_{x=x0}\\ &= -(1+\gamma)\,(G_N M/c^2 b)\,(\cos\theta_1 - \cos\theta_0).\end{aligned} \tag{38}$$

for the deflection angle $\Delta\varphi_{deflection}$, where $\theta_0$ ($\theta_1$) is the angle between the position vector of the light emitter (observer) and x-axis. For star light and close impact, i.e. $b << r_1$, we have

$$\Delta\varphi_{deflection} = -(1+\gamma)\,(G_N M/c^2 b)\,(\cos\theta_1 + 1). \tag{39}$$

If $\cos\theta_1 \approx 1$ and $\gamma = 1$, we obtain the usual formula of GR, i.e. $\Delta\varphi_{deflection} = -4\,(G_N M/c^2 b)$.

## 3. Solar System Ephemerides

Planetary ephemerides is a must for precision tests of relativistic gravity in the solar system and for the orbit design of spacecraft and missions. Before the advent of space age, the analytical theories developed by Le Verrier, Hill, Newcomb, and Clemens on planetary motion had sufficient accuracy to account for ongoing optical observations. With the Doppler radio tracking of spacecraft and the radio/laser ranging to planets/Moon, the required accuracy of planetary ephemerides increased tremendously. The accuracy of analytical theories became inadequate. The usage and development of numerical methods started.

Since the motion of planets and the moon are influenced by other planets/moon, to test relativistic gravity, one needs a complete solar-system ephemeris. To do this, one would start with the PPN equations of motion in an appropriate gauge for celestial bodies. Because the separation of planets/moon are large compared with their sizes, one could treat the planets/moon as point particles with suitable multipole moments. Such a set of PPN equations of motion is the post-Newtonian barycentric equations of motion as derived in Brumberg [73] from the post-Newtonian barycentric metric with PPN parameters $\beta$ and $\gamma$ for solar system bodies. The metric with the gauge parameter $\alpha$ (not to be confused with Eddington parameter $\alpha$ in the last section) and $\nu$ set to zero corresponding to a harmonic gauge adopted by the 2000 IAU resolution [74] is

$$ds^2 = [1 - 2\sum_i \frac{m_i}{r_i} + 2\beta(\sum_i \frac{m_i}{r_i})^2 + (4\beta - 2)\sum_i \frac{m_i}{r_i} \sum_{j \neq i} \frac{m_j}{r_{ij}}$$
$$- c^{-2} \sum_i \frac{m_i}{r_i}(2(\gamma+1)\dot{x}_i^2 - r_i \cdot \ddot{x}_i - \frac{1}{r_i^2}(r_i \cdot \dot{x}_i)^2) + \frac{m_1 R_1^2}{r_1^3} J_2(3(\frac{r_1 \cdot \hat{z}}{r_1})^2 - 1)]c^2 dt^2$$
$$+ 2c^{-1} \sum_i \frac{m_i}{r_i}((2\gamma + 2)\dot{x}_i) \cdot d\boldsymbol{x} c dt - [1 + 2\gamma \sum_i \frac{m_i}{r_i}](d\boldsymbol{x})^2 \qquad (40)$$

where $r_i = \boldsymbol{x} - \boldsymbol{x}_i, r_{ij} = \boldsymbol{x}_i - \boldsymbol{x}_j, m_i = GM_i/c^2$, and $M_i$'s are the masses of the celestial bodies with $M_1$ the solar mass [73]. $J_2$ is the quadrupole moment parameter of the Sun. $\hat{z}$ is the unit vector in the direction of solar angular momentum. The associated equations of motion of N-mass problem derived from the geodesic variational principle of this metric (the effect of solar quadrupole moment not yet added) are

$$\ddot{\boldsymbol{x}}_i = -\sum_{j \neq i} \frac{GM_j}{r_{ij}^3} \boldsymbol{r}_{ij} + \sum_{j \neq i} m_j (A_{ij} \boldsymbol{r}_{ij} + B_{ij} \dot{\boldsymbol{r}}_{ij}),$$

$$A_{ij} = \frac{\dot{\boldsymbol{x}}_i^2}{r_{ij}^3} - (\gamma+1)\frac{\dot{\boldsymbol{r}}_{ij}^2}{r_{ij}^3} + \frac{3}{2r_{ij}^5}(\boldsymbol{r}_{ij} \dot{\boldsymbol{x}}_j)^2 + G[(2\gamma+2\beta+1)M_i + (2\gamma+2\beta)M_j]\frac{1}{r_{ij}^4}$$

$$+ \sum_{k \neq i,j} GM_k[(2\gamma+2\beta)\frac{1}{r_{ij}^3 r_{ik}} + (2\beta-1)\frac{1}{r_{ij}^3 r_{jk}} + \frac{2(\gamma+1)}{r_{ij}^3 r_{jk}} - (2\gamma+\frac{3}{2})\frac{1}{r_{ik} r_{jk}^3} - \frac{1}{2r_{jk}^3}\frac{\boldsymbol{r}_{ij}\boldsymbol{r}_{ik}}{r_{ij}^3}],$$

$$B_{ij} = \frac{1}{r_{ij}^3}[(2\gamma+2)(\boldsymbol{r}_{ij}\dot{\boldsymbol{r}}_{ij}) + (\boldsymbol{r}_{ij}\dot{\boldsymbol{x}}_j)].$$

(41)

These equations can be used to build a computer-integrated planetary ephemeris framework. For a complete ephemeris, one needs to fit observational data to obtain solar and planetary parameters together with a set of initial conditions at a specific epoch; for a working ephemeris, one could simply adopt the parameters including planetary positions and velocities at some epoch from a complete fundamental ephemeris. For example, in our working CGC 1 ephemeris (CGC: Center for Gravitation and Cosmology) [75], we equations (41) with (40) (setting $\beta = \gamma = 1$ for general relativity, and $J_2 = 2 \times 10^{-7}$ for the Sun) for eight-planets plus Pluto, the Moon, the Sun and the 3 big asteroids - Ceres, Pallas and Vesta (14-body evolution); the positions and velocities at the epoch 2005.6.10 0:00 are taken from the DE403 ephemeris [76]. The evolution (can go forward or backward in time) is solved by using the 4$^{th}$-order Runge-Kutta method with the step size h =0.01 day. Since tilt of the axis of the solar quadrupole moment to the perpendicular of the elliptical plane is small (7°), in CGC 1 ephemeris, we have neglected this tilt. In CGC 2 ephemeris [77], we have added the perturbations of additional 489 asteroids. Such ephemerides can be used for mission orbit design/optimization and mission simulation. We used CGC 1 for orbit simulation and parameter determination for ASTROD. Using this ephemeris as a deterministic model and adding stochastic terms to simulate noise, we generated simulated ranging data and use Kalman filtering to determine the accuracies of fitted relativistic and solar-system parameters for 1050 days of the ASTROD mission [75]. This way, we simulated the accuracy achievable for the ASTROD mission concept. For a better evaluation of the accuracy of $\dot{G}/G$, we need also to monitor the masses of other asteroids. For this, we considered all known 492 asteroids with diameter greater than 65 km to obtain an improved ephemeris framework --- CGC 2, and calculated the perturbations due to these 492 asteroids on the ASTROD spacecraft [77]. More recently, we apply different variants of CGC 2 ephemeris framework to study the ASTROD I orbit design/optimization/simulation [78], the ASTROD-GW orbit design/optimization [79], and the numerical Time Delay Interferometry (TDI) for space gravitational-wave mission concepts ASTROD-GW [80], LISA [81] and eLISA [82].

At present, there are 3 series of complete fundamental ephemerides for the solar system – DE (Development Ephemerides) [83], INPOP (Intégrateur Numérique Planétaire de l'Observatoire de Paris) [84] and EPM (Ephemerides of Planets and Moon) [85]. The major common feature of these 3 series of ephemerides are the simultaneous numerical integration of the equations of motion of the eight planets plus Pluto, the Sun, the Moon, and the lunar physical libration using the post-Newtonian approximation of general relativity in a harmonic coordinate system. In addition, they take different numbers of asteroids and TNOs (trans-Neptunian objects) in the integration of ephemerides.

Let's illustrate with the EPM ephemerides [85]. Specifically, the basic dynamical model of EPM2011 is the post-Newtonian equations of motion of the Sun, the Moon, the 8 planets plus Pluto (now a TNO), and five largest asteroids with the following relatively weak gravitational effects taken into account:

(a) perturbations from the known 301 of the most massive asteroids;
(b) perturbations from other minor planets in the main asteroid belt, modeled by a homogeneous ring;
(c) perturbations from the known 21 largest TNOs;
(d) perturbations from the other trans-Neptunian planets, modeled by a homogeneous ring at a mean distance of 43 AU;
(e) perturbations from the solar oblateness ($2 \times 10^{-7}$).

The main data set for the 3 current ephemerides to fit comes from the astrometric observations of planets and spacecraft. For EPM2011 [85], it includes (i) Optical observations of the outer planets and their satellites made from 1913 to 2011 (57560 data points); (ii) Radar observations of Mercury, Venus, and Mars from 1961 to 1997 (58112 data points); (iii) Radio data provided by spacecraft from 1971 to 2010 (561998 data points).

From (41) with (40) or equivalent formulas and observations to obtain a complete ephemeris, one needs to fit for the mass parameters, relevant multipole moments, initial positions and initial velocities of the planets, the Moon and the Sun together with some other solar system bodies like the three largest asteroids − Ceres, Pallas and Vesta. Before 2009, in fitting the data for ephemerides, instead of the mass parameter $GM_{Sun}$ of the Sun, one could use the following relation to fit or adjust the astronomical unit au:

$$GM_{Sun}[m^3 s^{-2}] = k^2 \, au[m^3] / 86400^2 [s^2] \qquad (42)$$

with $k = 0.017\,202\,098\,95$ the Gaussian gravitational constant. The astronomical unit is a basic unit in astronomy and was supposed to be close to the mean Earth distance to the Sun. With development of the ranging observations in the solar system, it could be related precisely to the SI meter through the ephemeris fitting. Standish determined the au to be 149 597 870 697.4 m when worked out DE410 in 2003 [86]. Pitjeva [87] determined the au to be 149 597 870 696.0 m when worked out EPM2004. The difference of 1.4 m represents the realistic error in the determination of the au. In 2009, Pitjeva and Standish [88] proposed to the IAU Working group on Numerical Standards for fundamental Astronomy (NSFA) the masses of three largest asteroids (Ceres, Pallas and Vesta), the ratio of Moon's mass to Earth's mass, and the au from the fitting/adjustment of DE421 [89] and EPM2009 [90]. In this determination from ephemerides, the DE421 [89] value of au is 149 597 870 699.6 ± 0.15 m and the EPM 2009 [90] value of au is 149 597 870 696.6 ± 0.1 m with the quoted uncertainty formal uncertainty. They estimated the realistic error to be 3 m. From this result, they proposed to adopt the numerical value of the au in meter to be 149 597 870 700 (3) m. They also concluded that the numerical value of the au in meters is identical in both the TDB-based (TDB: Barycentric Dynamical Time) and the TCB-based (TCB: Barycentric Coordinate Time) systems of units if one uses the conversion proposed by Irwin and Fukushima [91], Brumberg and Groten [92], and Brumberg and Simon [93]. This value of au was accepted and included by the XXVIIth IAU General Assembly at Rio de Janeiro in 2009 as part of the IAU (2009) System of Astronomical Constants (Resolution B2) [94].

In the XXVIIIth IAU General Assembly at Beijing in 2012, the astronomical unit in meters is changed from a fitted value to a defining constant similar to speed of light: 1 au = 149 597 870 700 m. The abbreviation of the astronomical unit should be au (lower case). In the fitting of ephemeris, $GM_{Sun}$ should then be used instead of au. The current values from the ephemeris fitting are: DE430 [83], $GM_{Sun}$ = 132 712 440 042(10) km$^3$/s$^2$; INPOP13b,c [95], $GM_{Sun}$ = 132 712 440 044.5(0.2) km$^3$/s$^2$; EPM2014 [96], $GM_{Sun}$ = 132 712 440 053(1) km$^3$/s$^2$.

In the actual testing of relativistic theories of gravity, one fits additional PPN or relativistic parameters. In the next section, we compile these ephemeris tests together with other solar-system tests.

## 4. Solar System Tests

For last fifty years, we have seen great advances in the dynamical testing of relativistic gravity. This is largely due to interplanetary radio ranging/tracking and lunar laser ranging (LLR). Interplanetary radio ranging and tracking provided more stimuli and progresses at first. However, with improved accuracy of 2 mm from 20-30 cm and long-accumulation of observation data, lunar laser ranging reaches similar accuracy in determining relativistic parameters as compared to interplanetary radio ranging despite that in LLR the relativistic effects are weaker. Table 2 gives such a comparison.

*Relativistic perihelion advance and the solar quadrupole moment.* In the PPN equations of motion (41) with two PPN parameters $\beta$ and $\gamma$ and with the effect of solar quadrupole moment added, the solar contribution to the secular planetary perihelion advances is given by the well-known formula (see, e.g. [48], p.1116):

$$\Delta\phi_0 = [(2 - \beta + 2\gamma)/3] [6\pi GM/(c^2 a(1 - e^2))] + J_2 [3\pi R^2/(a^2(1 - e^2)^2)], \tag{43}$$

where $a$ and $e$ are the semimajor axis and the eccentricity of the planet orbit. If Sun is uniformly rotating throughout, $J_2$ would be about $2 \times 10^{-7}$ and its magnitude amounts to about 0.05 % of the general relativistic perihelion advance for Mercury. To measure or separate the relativistic term (the first term), one needs to know or measure the solar quadrupole parameter. There are 3 ways to measure the solar quadrupole parameter: (i) through the solar oblateness measurement based on brightness of solar surface; (ii) through the helioseismology measurement; (iii) through the measurement of perihelion advance of different planets and asteroids. Although the solar oblateness measurement up to 1980's might imply large solar quadrupole parameter [97], the determination of internal rotation of the Sun through measurement of rotation-induced frequency splitting in the observed solar surface acoustic power spectrum about the same period of time gave the value $J_2 = (1.7 \pm 0.4) \times 10^{-7}$, rather close to the value for a uniformly rotating Sun [98].

In the 12$^{th}$ International conference on General Relativity and Gravitation at Boulder in 1989, Shapiro [99] reported the cumulative measurement of the relativistic Mercury perihelion advance to be (42".98 ± 0.04)/century assuming the solar quadrupole parameter $J_2$ to be about $2 \times 10^{-7}$; this gives $\beta = 1.000 \pm 0.003$.

In the 1990's, the solar quadrupole moment issue basically settled: (i) The solar oblateness $\varepsilon$ measured in a balloon flight of the Solar Disk Sextant (SDS) on 1992 September 30 [100] is $(8.63 \pm 0.88) \times 10^{-6}$ with the inferred $J_2 = 3 \pm 6 \times 10^{-7}$ (in agreement with the measured and inferred values $\varepsilon = (9.6 \pm 6.5) \times 10^{-6}$ and $J_2 = 10 \pm 43 \times 10^{-7}$ of Hill and Stebbins [101] in 1975). A subsequent analysis [102] based on SDS balloon flight data both in 1992 and 1994 combined with solar surface angular rotation data gives the solar quadrupole moment parameter $J_2 = 1.8 \times 10^{-7}$ and the solar octopole moment parameter $J_4 = 9.8 \times 10^{-7}$. (ii) The high quality helioseismological data obtained from the Solar Heliospheric Satellite (SoHO) and from the Global Oscillations Network Group (GONG) had made a much better determination of solar internal structure and solar differential rotation possible; this in turn led to a good determination of solar quadrupole moment and solar angular momentum. Pijpers [103] did an analysis and obtained $J_2 = (2.14 \pm 0.09) \times 10^{-7}$ from the GONG data and $J_2 = (2.23 \pm 0.09) \times 10^{-7}$ from the SoHO/MDI data, with an error-weighted mean $J_2 = (2.18 \pm 0.06) \times 10^{-7}$. Godier and Roselot [104] used a differential rotation model established from helioseismological data and integrated $J_2$ from core to the surface to obtain a slightly lower value of $J_2 = 1.60 \times 10^{-7}$.

The results of space-borne measurements of solar oblateness from 1997 to 2011 are basically giving consistent numbers as summarized by Meftah *et al*. in 2015 [105]: SoHO/MDI by Emilio *et al*. in 2007 [106] with the oblateness (the solar equator-to-pole radius difference) $\Delta r = 8.7 \pm 2.8$ mas using 676.78 nm ($\lambda$) observation (2007), RHESSI/SAS by Fivian *et al*. in 2008 [107] with $\Delta r = 8.01 \pm 0.14$ mas using 670.0 nm observation in 2004, SDO/HMI by Kuhn *et al*. (2012) [108] with $\Delta r = 7.2 \pm 0.49$ mas using 617.3 nm observation (2011 – 2012), *Picard*/SODISM by Irbah *et al*. (2014) [109] with $\Delta r = 8.4 \pm 0.3$ mas using 535.7 nm observation (2011), *Picard*/SODISM by Meftah *et al*. (2015) [105] $\Delta r = 7.86 \pm 0.32$ mas using 782.2 nm observation (2010 – 2011). It is to be noted that $\Delta r = 8$ mas corresponds to $J_2 = 1.60 \times 10^{-7}$.

The third way (iii) to measure the solar quadrupole moment is through its gravitational field generated. This will be discussed in the following together with the ephemeris fitting.

*Test of relativistic gravity through ephemeris fitting.* As planetary ephemerides became more and more precise, Anderson *et al*. [110] in 2002, used JPL archive of planetary positional data and DE ephemerides fitting method to solve for all the conventional parameters in the DE ephemeris, plus four more parameters $\beta$, $\gamma$, $J_2$ and $\dot{G}/G$ specific to tests of relativistic gravity. In fitting the data, they weighted the separate data sets, except four data sets for the Mars, such that the assumed standard error for each data set is equal to the rms residual for that particular set after the fit. For Mars, they use a standard error equal to 5 times the rms residual for each of the four data sets -- orbit data from Mariner 9 (1971-1972), lander data from Viking (1976-1982), orbit data from Mars global Surveyor (1998-2000) and Lander data from Pathfinder (1997), to compensate for systematic error from asteroids perturbations. This way they interpreted their resulting parameter values after fit as realistic errors instead of formal errors. The results of their fitting [110] are $\beta = 0.9990 \pm 0.0012$, $\gamma = 0.9985 \pm 0.0021$, $J_2 = (2.3 \pm 5.2) \times 10^{-7}$, and $\dot{G}/G = \pm(1.1\text{-}1.8)\times 10^{-12}$/yr (the $\dot{G}/G$ value is the same as their previous result in [111]).

As an application of the developing EPM ephemerides, Pitjeva [87] (EPM2004 fitting) in 2004 obtained a determination of $\beta$ and $\gamma$ simultaneously with estimations for the solar quadrupole parameter and the possible variability of the gravitational constant: $\beta = 1.0000 \pm 0.0001$, $\gamma = 0.9999 \pm 0.0001$, $J_2 = (1.9 \pm 0.3) \times 10^{-7}$ and $\dot{G}/G = (1 \pm 5)\times 10^{-14}$/yr. In working out INPOP2010a planetary ephemeris, Fienga *et al*. [112] tested relativistic gravity by fitting $\beta$ or $\gamma$, and obtained: $\beta = 0.999959\pm0.000078$; $\gamma = 1.000038 \pm 0.000081$; $J_2 = (2.4 \pm 0.25) \times 10^{-7}$. Pitjeva in working out EPM2011 ephemerides in 2013 [85] obtained $\beta = 0.99998\pm0.00003$, $\gamma = 1.00004 \pm 0.00006$, $J_2 = (2.0 \pm 0.2) \times 10^{-7}$ and $[d(GM_{Sun})/dt]/GM_{Sun} = (-5.0 \pm 4.1)\times10^{-14}$/yr. Verma *et al*. [113] included the radio ranging observations of MESSENGER, improved our knowledge of the orbit of Mercury, obtained INPOP13a ephemeris, and used it to perform tests of relativistic gravity. Their estimations of parameters are: $\beta = 1.000002\pm0.000025$; $\gamma = 0.999997 \pm 0.000025$ ($\beta = 1$ fixed); $J_2 = (2.4 \pm 0.2) \times 10^{-7}$.

Fienga, Laskar, Manche and Gastineau [114] added supplementary range tracking data obtained from the analysis of the MESSENGER spacecraft from 2011 to 2014 and included in their INPOP15a planetary ephemerides the new JPL datasets [115] obtained after the new analysis of Cassini tracking data from 2004 to 2014. They use INPOP15a to estimate possible supplementary advances of perihelia for Mercury and Saturn to tests GR and presented their results in the 14th Marcel Grossmann meeting [115]. The results are basically consistent with previous analysis; no violations of GR are found.

Using analytic and numerical methods, Anderson, Gross, Nordtvedt and Turyshev [116] demonstrated that Earth-Mars ranging could provide a useful estimate of the strong equivalence principle (SEP) parameter $\eta$. For Mars ranging measurements with an accuracy of $\sigma$ meters for 10 years, the expected accuracy for the Nordtvedt SEP parameter $\eta$ would be of order $(1-12) \times 10^{-4}\sigma$ according to [116]. The strong equivalence principle for the Earth-Mars-Sun-Jupiter system is probably already tested implicitly in the ephemeris fit. It remains to separate the effect of strong equivalence principle violation in the fit.

*Time variability of the gravitational constant and mass loss from the Sun.* The solar-system dynamics could measure the possible time variability of the gravitational constant and the mass change of the Sun when the precision becomes good. If the gravitational constant does not change or its change is measured in another way, the mass loss (change) from the Sun can be measured dynamically. We advocate this potential during the 1990's when we propose the concept of ASTROD [117]. The electromagnetic radiation of the Sun carries $6.8 \times 10^{-14}$ fractional mass from the Sun each year. This is the largest mass change of the Sun. Other mass change mechanisms give similar fractional change but of smaller magnitude. In order to separate the time variability of the gravitational constant, estimation of the mass loss from Sun and the mass accretion into Sun is needed. Pitjeva and Pitjev [118] made an estimate of the mass of celestial bodies falling into Sun (mainly comets) and gave the following annual upper limit:

$$M_{comet}/M_{Sun} < 3.2 \times 10^{-14}. \tag{44}$$

Combined with the estimate of annual solar wind loss of $(2 - 3) \times 10^{-14}$ $M_{Sun}$ per year (See [118] for references), Pitjeva and Pitjev gave the following bounds on the annual mass loss $M_{loss}$ of Sun:

$$-9.8 \times 10^{-14} < M_{loss}/M_{Sun} < -3.6 \times 10^{-14}. \tag{45}$$

The fitted value of

$$[d(GM_{Sun})/dt]/GM_{Sun} = (-5.0 \pm 4.1) \times 10^{-14}/\text{yr}, \quad (3\sigma) \tag{46}$$

from EPM [85, 118] led to the following relation [118] with 95% confidence (2σ) level

$$-7.8 \times 10^{-14} \text{ /yr} < (\dot{G}/G + M_{\text{Sun}}\text{-dot}/M_{\text{Sun}}) < -2.3 \times 10^{-14} \text{ /yr, (2σ)}. \tag{47}$$

Eq. (47) together with (45) gave bound on $\dot{G}/G$ [118] as

$$-4.2 \times 10^{-14} \text{ /yr} < \dot{G}/G < +7.5 \times 10^{-14} \text{ /yr}. \tag{48}$$

More recently, Fienga *et al.* [119] used Monte Carlo simulations to find constraints on the possible variation of the gravitational constant. They deduced the values of $\dot{G}/G$ considering a fixed value for annual mass loss of the Sun (including radiation and solar winds):

$$M_{\text{loss}}/M_{\text{Sun}} = (5.5 \pm 1.5) \times 10^{-14}, \tag{49}$$

extracted from solar physics measurements and variations of $M_{\text{loss}}/M_{\text{Sun}}$ during the 11-year solar cycle of Pinto *et al*. [120]. The values of $\dot{G}/G$ are typically within $\pm 10 \times 10^{-14}$.

Solar-system dynamics also constrains dark energy models. For interested readers, please see Refs. [121, 122].

*Light/radio wave deflection, Shapiro time delay and constraint on the Eddington parameter γ*. As we have seen in Sec. 1, gravitational light deflection is one of three classical tests of GR. Before the ephemeris determination of the Eddington (light deflection) parameter *γ*, the VLBI (Very Long Baseline Interferometry) measurement of the gravitational deflection of radio waves by the Sun from astrophysical radio sources had been an important method. The accuracy of observation had been improved to $1.7 \times 10^{-3}$ for *γ* (Robertson *et al.* [123], Lebach *et al.* [124] and references therein) in 1995. Analysis using VLBI data from 1979-1999 improved the result by about four times to $0.99983 \pm 0.00045$ (Shapiro *et al*. [125]). Fomalont, Kopeikin, Lanyi and Benson [126] used the Very Long Baseline Array (VLBA) at 43, 23 and 15 GHz to measure the solar gravitational deflection of radio waves among four radio sources during an 18-day period in October 2005 and determined the Eddington parameter *γ* to be $0.9998 \pm 0.0003$. Fomalont and Kopeikin [127, 128] measured the effect of retardation of gravity by the field of moving Jupiter via VLBI observation of light bending from a quasar.

In 2003, Bertotti, Iess and Tortora [129] reported a measurement of the frequency shift of radio photons due to relativistic Shapiro time-delay effect from the Cassini spacecraft as they passed near the Sun during the June 2002 solar conjunction. From this measurement, they determined *γ* to be $1.000021 \pm 0.000023$.

With the Hipparcos mission, very accurate measurements of star positions at various elongations from the Sun were accumulated [130]. Most of the measurements were at elongations greater than 47° from the Sun. At these angles, the relativistic light deflections are typically a few mas; it is 4.07 mas at right angles to the solar direction for an observer at 1 AU from the Sun according to GR. In the Hipparcos measurements, each abscissa on a reference great-circle has a typical precision of 3 mas for a star with 8-9 mag. There are about 3.5 million abscissae generated, and the precision in angle or similar parameter determination is in the range. Frœschlé, Mignard and Arenou [131] analyzed these Hipparcos data and determined the light deflection parameter *γ* to be $0.997 \pm 0.003$. This result demonstrated the power of precision optical astrometry.

Gaia (Global Astrometric Interferometer for Astrophysics) [132] is an ambitious astrometric mission aiming at the broadest possible astrophysical exploitation of optical interferometry using a modest baseline length (~3m). Gaia, launched on 19 December 2013 by Arianespace using a Soyuz ST-B/Fregat-MT rocket flying from Kourou in French Guiana in a Lissajous orbit around the Sun–Earth $L_2$ Lagrangian point, is charting a three-dimensional map of our Galaxy, the Milky Way, in the process revealing the composition, formation and evolution of the Galaxy. Operating in the depths of space, far beyond the Moon's orbit, ESA's Gaia spacecraft had completed two years of a planned five-year survey of the sky on 16 August 2016. Data Release 1 (Gaia DR1) [133] was already public and contained astrometric results for more than 1 billion stars brighter than magnitude 20.7 based on observations collected by the Gaia satellite during the first 14 months of its operational phase. Gaia has already provided unprecedented positional and radial velocity measurements with the accuracies needed to produce a stereoscopic and kinematic census of about one billion stars in our Galaxy and throughout the Local Group. This amounts to about 1 per cent of the Galactic stellar population. To increase the weight of measuring the relativistic light deflection parameter γ, Gaia observes at elongations greater than 35° (as compared to essentially 47° for Hipparcos) from the Sun. A simulation shows that GAIA could measure $γ$ to $1×10^{-5} – 2 ×10^{-7}$ accuracy [133].

*Lunar Laser Ranging (LLR) Tests of relativistic gravity.* In the last column of Table 2, the values come from LLR observations [135-140]. Ref. [136] gave the results as of 1996. In [136], Williams *et al.* used a total of 15 553 LLR normal-point data in the period of March 1970 to April 2004 from Observatoire de la Côte d'Azur, McDonald Observatory and Haleakala Observatory in their determination. Each normal point comprises from 3 to about 100 photons. The weighted rms scatter after their fits for the ten-year ranges from 1994 to 2004 is about 2 cm (about $5 × 10^{−11}$ of range). Müller *et al.* wrote a comprehensive chapter on "Lunar Laser Ranging and Relativity" and summarize their work on the LLR tests of the relativistic gravity [138]. From Table 2, we can see clearly that the LLR tests of relativistic gravity have the same level of precision as the radio solar-system tests. Constraints on intermediate range force is from LLR [138] and from the Mars perihelion precession (Iorio [141] and references therein) are compiled in the last row.

LLR also constrains dark energy models. For interested readers, please see [142] and references therein.

*Frame Dragging Effects.* In 1918, Lense and Thirring [144] predicted that a rotating body drags the local inertial frames of reference around it in general relativity. In 1960, L. I. Schiff [145] showed that in general relativity the spin axis of a gyroscope orbiting around Earth would undergo both geodetic drift in the orbit plane due to motion through the space-time curved by the Earth's mass and frame-dragging due to the Earth's rotation with respect to a distant inertial frame. The dragging of gyro's spin axis is sometimes called the Schiff effect while both spin axis dragging and orbiting axis dragging can be grouped as Lense-Thirring frame-dragging effects. In 2004, Ciufolini and Pavlis [146] reported a measurement of the Lense-Thirring effect on the two Earth satellites, LAGEOS and LAGEOS2; it is 0.99 ± 0.10 of the value predicted by general relativity. In the same year, Gravity Probe B (a space mission to test general relativity using cryogenic gyroscopes in orbit [147]) was launched in April [148]; their final results are a geodetic drift rate of −6,601.8±18.3 mas/yr and a frame-dragging drift rate of −37.2±7.2 mas/yr, to be compared with the GR predictions of −6,606.1 mas/yr and −39.2 mas/yr, respectively; i.e., GP-B [148] provides independent measurements of the geodetic and frame-dragging effects at an accuracy of 0.28% and 19%, respectively. GP-B experiment has also verified the weak equivalence principle for macroscopic rotating bodies to ultra-precision [149]. Recently, Ciufolini *et al.* [150] have used about 3.5 years of laser-range observations of the LARES, LAGEOS, and LAGEOS2 satellites together with the Earth gravity field model GGM05S produced by the space geodesy mission GRACE to measure the Earth's dragging of inertial frames to be 0.994 ± 0.002 ± 0.05 of the general relativity value with 0.002 the 1-$\sigma$ formal error and 0.05 their preliminary estimate of systematic error.

Table 2. Relativity-parameter determination from interplanetary radio ranging/tracking and from lunar/satellite laser ranging.

| Parameter | Meaning | Value from Solar System Determinations and from Gravity Probe B | Value from Lunar/Satellite Laser Ranging |
|---|---|---|---|
| $\beta$ | PPN [55] Nonlinear Gravity | 1.000±0.003 [99] (Perihelion shift with $J_2$(Sun) = $10^{-7}$ assumed)<br>0.9990±0.0012 [110] (Solar-system tests with $J_2$(Sun) = (2.3±5.2)×$10^{-7}$ fitted)<br>1.0000±0.0001 [87] (EPM2004 fit)<br>0.999959±0.000078 [112] (INPOP10a fit)<br>0.99998±0.00003 [85] (EPM2011 fit) | 1.003±0.005 [135]<br>1.00012±0.0011 [136, 129] (with Cassini $\gamma$)<br>1.00017±0.00015 [138]<br>1.00006±0.00011 [138] (from $\eta$) |
| $\gamma$ | PPN [55] Space Curvature | 1.000±0.002 [99] (Viking ranging time delay)<br>0.9985±0.0021 [110] (Solar-system tests)<br>1.000021±0.000023 [129](Cassini S/C ranging)<br>0.9999±0.0001 [87] (EPM2004 fit)<br>0.9998±0.0003 [126] (VLBI deflection)<br>1.000038±0.000081 [112] (INPOP10a fit)<br>1.00004±0.00006 [85] (EPM2011 fit) | 1.000±0.005 [135] |
| $K_{gp}$ | Geodetic Precession | 0.99935±0.0028 [148] (Gravity Probe B) | 0.997±0.007 [135]<br>0.9981±0.0064 [136]<br>0.997±0.005 [138] |
| $K_{L-T}$ | Lense-Thirring Effect | 0.95±0.19 [148] (Gravity Probe B) | 0.994±(0.1-0.3) [146] (LAGEOS)<br>0.994±0.002± 0.05 [150] (LAGEOS & LARES) |
| E ($\eta$) | Strong Equivalence Principle (Nordtvedt parameter) | The strong equivalence principle for the Earth-Mars-Sun-Jupiter system [116] is probably already tested implicitly in the ephemeris fit. It remains to separate the effect of strong equivalence principle violation. | (3.2±4.6)×$10^{-13}$ [135]<br>(-2.0±2.0)×$10^{-13}$ [136,143]<br>(−0.8 ± 1.3) × $10^{-13}$ [137]<br>(0.9±1.9) × $10^{-13}$ [138]<br>($\eta$ = (1±3)×$10^{-4}$[139]) |
| $\dot{G}/G$ | Temporal Change in $G$ | ±(1.1-1.8)×$10^{-12}$/yr [111] (Solar-System Tests)<br>(1 ± 5)×$10^{-14}$/yr. [87] (EPM2004 fitting)<br>(−4.2 to 7.5)× $10^{-14}$/yr [118] (Planets & S/C observations with solar mass loss estimate)<br>± 10 ×$10^{-14}$/yr [119] (INPOP & Monte Carlo with solar mass loss estimate) | (1±8) × $10^{-12}$/yr [135]<br>(4±9)× $10^{-13}$/yr [136]<br>(14±15)× $10^{-14}$/yr [138] |
| $\ddot{G}/G$ | Temporal Change in $\dot{G}$ | | (4±5) × $10^{-15}$ yr$^{-2}$ [140] |
| $\alpha_{Yukawa}$ | Intermediate range force | $\alpha_{\lambda=1.5\ au}$ = (2±13)×$10^{-10}$ [141] (perihelion of Mars) | $\alpha_{\lambda=380\ 000\ km}$ = (−0.6±1.8)× $10^{-11}$ [138] |

## 5. Outlook – On Going and Next-Generation Tests

In the early days, astronomical observations of the solar system provided the basis for developing gravitation theories. Gravitation theories provide the scientific basis of space exploration of Earth and the entire solar system. The advent of space age and solar-system exploration required the range measurements in the solar-system that made possible the creation of high-accuracy planetary and lunar ephemerides. These ephemerides in turn provide dynamical positioning atlases for the solar-system exploration and the precision tests of relativistic gravitational theories. As we have seen in the last section, ephemeris fitting for gravitational parameters in relativistic gravitational theories is playing more and more important role in the experimental tests. Experimentally, the improvement depends on the technological advance of radio ranging/Doppler tracking and laser ranging/tracking of spacecraft and celestial bodies in the solar system.

In Table 2, we have seen that lunar laser ranging reaches similar accuracy in determining relativistic parameters as compared to interplanetary radio ranging despite that in LLR the relativistic effects are weaker. The main reason is that the resolution depends on wavelength. Optical wavelength is 4 orders of magnitude shorter than microwave wavelength. The most precise radio Doppler tracking experiment is Cassini radio wave retardation measurement [129]. Cassini multilink radio system consists of a sophisticated multilink radio system that simultaneously receives two uplink signals at frequencies of X and Ka bands and transmits three downlink signals with X-band coherent with the X-band uplink, Ka-band coherent with the X-band uplink, and Ka-band coherent with the Ka-band uplink. X band is a standard deep space communication frequency band about 8.4 GHz; Ka band is another deep space communication frequency band about 32 GHz. The wavelength of Ka band microwave is about 1 cm. The reason to use multilink system is to measure and subtract the plasma dispersion which is proportional to the wavelength square. For laser optical ranging, a typical wavelength is about 1 μm. There is a four-order difference in wavelength. For laser ranging, the plasma effect is eight-order smaller; in the interplanetary space the subtraction is not needed. If one link is on Earth, subtraction of extra optical path length by two-wavelength observation or other means is still needed. With four-order improvement in ranging, monitoring the non-inertial spacecraft motion is required. One way is to use drag-free technology. LISA Pathfinder launched on 3 December, 2015 has successfully tested and demonstrated the drag-free technology to satisfy not just the requirement of LISA Pathfinder, but also basically the drag-free requirement of LISA gravitational-wave space mission concept [151]. The drag-free technology is ripe for relativistic missions in the solar-system. Hence, we envisage a 3-4 orders of improvement in testing the relativistic gravity and the solar-system dynamics, say, in the next 25 years or so. This improvement is for all relativistic parameters. In the following we give an outlook of improvements on the Eddington parameter $\gamma$ for various ongoing/proposed experiments. Table 3 lists the aimed accuracy of such experiments. Some motivations for determining $\gamma$ precisely to $10^{-5} - 10^{-9}$ are given in [152, 153].

First, as we have discussed in Sec. 4, Gaia Data Release 1 (Gaia DR1) [133] has already become public and contained astrometric results for more than 1 billion stars brighter than magnitude 20.7 based on observations collected by the Gaia satellite during the first 14 months of its operational phase. With expected 4-year observation period, a simulation shows that GAIA could measure $\gamma$ to $1\times10^{-5} - 2\times10^{-7}$ accuracy [133]. This is listed as the second row in Table 3.

BepiColombo is a joint mission to Mercury [154] between ESA and the Japan Aerospace Exploration Agency (JAXA), executed under ESA leadership. The mission comprises two spacecraft: The Mercury Planetary Orbiter (MPO) and the Mercury Magnetospheric Orbiter (MMO). It will set off in 2018 on a journey to the smallest and least explored terrestrial planet in our Solar System. When it arrives at Mercury in late 2024, it will endure temperatures in excess of 350 °C and gather data during its 1-year nominal mission, with a possible 1-year extension. Milani, Vokrouhlicky, Villani, Bonanno and Rossi [155] have simulated the radio science of this mission: "While determining its orbit around Mercury, it will be possible to indirectly observe the motion of its center of mass, with an accuracy several orders of magnitude better than what is possible by radar ranging to the planet's surface. This is an opportunity to conduct a relativity experiment which will be a modern version of the traditional tests of general relativity, based upon Mercury's perihelion advance and the relativistic light propagation near the Sun." They predict that the determination of $\gamma$ can reach $2 \times 10^{-6}$.

ASTROD I is envisaged as the first in a series of ASTROD missions [78, 156-159]. ASTROD I mission concept is to use one spacecraft carrying a telescope, four lasers, two event timers and a clock with a Venus swing-by orbit. Two-way, two-wavelength laser pulse ranging will be used between the spacecraft in a solar orbit and deep space laser stations on Earth, to achieve the ASTROD I goals of testing GR with an improvement in sensitivity of over 3 orders of magnitude, improving our understanding of gravity and aiding the development of a new quantum gravity theory; to measure key solar system parameters with increased accuracy; and to measure the time rate of change of the gravitational constant with improvement. Using the achieved accuracy of 3 ps in laser pulse timing and the demonstrated LISA Pathfinder drag-free capability, a simulation showed that accuracy of the determination of $\gamma$ will reach $3 \times 10^{-8}$.

The general concept of ASTROD (Astrodynamical Space Test of Relativity using Optical Devices) is to have a constellation of drag-free spacecraft navigate through the solar system and range with one another using optical devices to map the solar-system gravitational field, to measure related solar-system parameters, to test relativistic gravity, to observe solar g-mode oscillations, and to detect gravitational waves. A baseline implementation of ASTROD, also called ASTROD, is to have two spacecraft in separate solar orbits (one in inner solar orbit, the other in outer solar orbit), each carrying a payload of a proof mass, two telescopes, two 1- 2 W lasers, a clock and a drag-free system, together with a similar spacecraft near Earth around one of the Lagrange points L1/L2. The three spacecraft range coherently with one another using lasers to map solar-system gravity, to test relativistic gravity, to observe solar g-mode oscillations, and to detect gravitational waves. Since it will be after ASTROD I, we assume 1 ps timing accuracy and 10 times better drag-free performance than what LISA Pathfinder achieved. With these requirements, the accuracy of the determination of $\gamma$ will reach $1 \times 10^{-9}$ in 3.5 years [160].

Super-ASTROD [161], Odyssey [162], SAGAS (Search for Anomalous Gravitation using Atomic Sensors) [163] and OSS (Outer Solar System) [164] are four mission concept to test fundamental physics and to explore outer solar system.

Solar System Odyssey [162] is designed to perform a comprehensive set of gravitational tests in the Solar System. The mission has four major scientific objectives: (1) significantly improve the accuracy of deep space gravity test; (2) investigate planetary flybys; (3) improve the current accuracy of the measurements of the Eddington parameter; (4) map the gravity field in outer regions of the Solar System. For improving the current accuracy of the measurement of the Eddington parameter $\gamma$, Odyssey proposes to use an improved multi-frequency radio link of the Cassini type together with a precision accelerometer and a possible laser tracking option and aims at measuring $\gamma$ at an accuracy of $10^{-7}$.

SAGAS [163] aims at flying highly sensitive atomic sensors (optical clock, cold atom accelerometer, optical link) on a Solar System escape trajectory. It also aims at measuring $\gamma$ at an accuracy of $1\text{-}2 \times 10^{-7}$.

OSS [164] is an outer solar system exploration mission concept. The OSS probe would carry instruments allowing precise tracking of the spacecraft during the cruise. It would facilitate improved tests of the laws of gravity in deep space. A largely improved accuracy can be attained with the up-scaling option of a laser ranging equipment onboard to measure the parameter $\gamma$ at the $10^{-7}$ level.

Super-ASTROD [161] is a mission concept with 4 spacecraft in 5 AU orbits together with an Earth-Sun L1/L2 spacecraft ranging optically with one another to probe primordial gravitational waves with frequencies 0.1 μHz - 1 mHz, to test fundamental laws of spacetime and to map the outer-solar-system mass distribution and dynamics. With larger orbits, the main goal of Super-ASTROD in test relativistic gravity is not to improve on PPN (Parametrized Post-Newtonian) parameters over ASTROD I / ASTROD, instead it is to test cosmological theories which give larger modifications from general relativity for larger orbits. However, with same or better ranging capability than ASTROD I / ASTROD, the accuracy of its determination of $\gamma$ will be better than $1 \times 10^{-8}$.

All four mission concepts explore gravity at deep space to bridge the gap between inner solar-system tests and cosmological tests. They are most relevant to the detection/constraint of dark matter and dark energy, and to the tests of MOND models and the dark energy dynamical models.

Table 3. Aimed accuracy of PPN space parameter $\gamma$ for various ongoing/proposed experiments.

| Ongoing/Proposed experiment | Aimed accuracy of $\gamma$ | Type of experiment |
|---|---|---|
| GAIA [132-134] | $1\times10^{-5} - 2\times10^{-7}$ | deflection |
| Bepi-Colombo [154, 155] | $2\times10^{-6}$ | retardation |
| ASTROD I [116] | $3\times10^{-8}$ | retardation |
| ASTROD [118] | $1\times10^{-9}$ | retardation |
| Super-ASTROD [161] | $1\times10^{-8}$ | retardation |
| Odyssey [162] | $1\times10^{-7}$ | retardation |
| SAGAS [163] | $1\times10^{-7}$ | retardation |
| OSS [164] | $1\times10^{-7}$ | retardation |

Since we had another review on "Equivalence principles, spacetime structure and the cosmic connection" [17] early this year, we did not discuss space missions for testing (weak) equivalence principles. Here we just mention in passing that Microscope (MICROSCOPE: MICRO-Satellite à trainée Compensée pour l'Observation du Principe d'Équivalence) [165, 166] has been in orbit since 26 April, 2016 with the aim of improving the test accuracy by 2 orders of magnitude than any of the ground-based weak-equivalence-principle experiment, and is performing functional tests successfully [166].

With increasing outreach and precision of observations, astrophysics and cosmology became increasingly important for developing gravitation theories; notably the precise timing of pulses from pulsars [44] and various cosmological tests [17].

During last 157 years, the precisions of laboratory and space experiments, and the precisions of astrophysical and cosmological observations on the tests of relativistic gravity have improved by 3-4 orders of magnitude. The advent of space age has stimulated the development of numerical ephemerides. Doppler and ranging observations from various space missions drive the ephemerides to ever increasing precision. For the last decade, we have seen great progress in various aspects of testing relativistic gravity in the solar system. Systematic modelling and ephemeris fitting of all the observational data becomes standard. The pending testing of relativistic gravity to better than $10^{-5}$-$10^{-6}$ precision requires the development of 2PN (post-post-Newtonian) numerical ephemerides. In the next 25 years, we envisage another 3-4 order improvement in all directions of tests of relativistic gravity. These will give enhanced interest and development both in experimental and theoretical aspects of gravity, and may lead to answers to some profound questions of gravity and the cosmos.

Gravitation is clearly empirical. As precision is increased by orders of magnitude, we are in a position to explore deeper into the origin of gravitation. The current and coming generations hold such promises.

## Acknowledgements

I am grateful to Jürgen Müller and Franz Hofmann for helpful discussions on the LLR tests of relativistic gravity. I would also like to thank Science and Technology Commission of Shanghai Municipality (STCSM-14140502500) and Ministry of Science and Technology of China (MOST-2013YQ150829, MOST-2016YFF0101900) for supporting this work in part.## References